\documentclass[fleqn,twoside]{article}
\usepackage{espcrc2}
\usepackage{epsfig}

\usepackage{graphicx}
\usepackage[figuresright]{rotating}

\newcommand{\AmS}{{\protect\the\textfont2
  A\kern-.1667em\lower.5ex\hbox{M}\kern-.125emS}}

\newcommand{\be}{\begin{equation}}
\newcommand{\ee}{\end{equation}}
\newcommand{\ba}{\begin{eqnarray}}
\newcommand{\ea}{\end{eqnarray}}

\newcommand{\nn}{\nonumber}

\newcommand{\op}{{\cal O}}

\newcommand{\fig}{Fig.~}

\title{The phase diagram of $N_f=3$ QCD for small baryon densities}

\author{Ph.~de Forcrand\address{ETH, CH-8093 Z\"urich, Switzerland and
CERN Theory Division, CH-1211 Geneva 23, Switzerland}
        and
        O.~Philipsen\address{Center for Theoretical Physics, MIT,
Cambridge, MA 02139-4307, USA}
\thanks{Talk given by O.~P.}}
       
\begin{document}

\begin{abstract}
We demonstrate how to locate the critical endpoint of the QCD phase transition
by means of simulations at imaginary $\mu$. For the three flavor theory, 
we present numerical results for the
pseudo-critical line as a function
of chemical potential and bare quark mass, as well as the bare quark mass dependence of
the endpoint.
\vspace{1pc}
\end{abstract}

\maketitle

\section{INTRODUCTION}

As of last year's lattice conference, there were three numerical methods
\cite{fk2}-\cite{fp1} capable of simulating the small $\mu/T$ regime of QCD
($\mu_B=3\mu$)
with mutually agreeing results on the pseudo-critical line $T_c(\mu_B)$ \cite{lp}. 
However, only one of them \cite{fk2} has studied 2+1 flavor QCD
and obtained a prediction for the location of the critical endpoint of the 
deconfinement transition. Here we summarize new results for three degenerate
flavors of standard staggered fermions, obtained by simulations at imaginary 
chemical potential followed by analytic continuation \cite{fp2}, as previously
applied to the two flavor case \cite{fp1} and other observables \cite{immu}.
For a discussion of the QCD symmetries at imaginary $\mu$ and the general strategy
of obtaining the critical line we refer to \cite{fp1}. Here we pay special
attention
to the location of the critical endpoint as a function of quark mass $m$.

In the three-dimensional parameter space $(T,\mu,m)$ the pseudo-critical temperature
represents a surface $T_c(m,\mu)$. On this surface, the line of critical endpoints
$T^*(\mu)=T_c(m_c(\mu),\mu)$ separates regions of first order transitions from crossover
behavior.
For three standard staggered flavors and at zero density, 
we know the ``chiral critical point'' $m_c(\mu=0)$ \cite{mc0}, 
above which the first order deconfinement transition turns into a crossover.
It belongs to the universality class of the 3d Ising model.
\begin{figure}[tbh]
\centerline{\epsfxsize=7cm\hspace*{0cm}\epsfbox{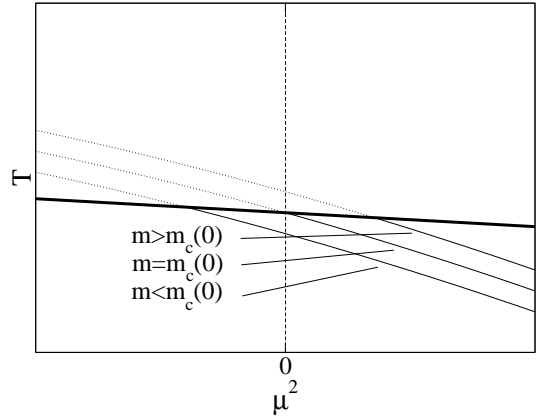}}
\vspace*{-0.8cm}
\caption[a]{\label{schem}
Critical lines in the $(T,\mu^2)$-plane for different quark masses,
dotted lines denote crossover.
The bold curve represents $T^*(\mu^2)$.}
\end{figure}
For this quark mass, the transition in the $(\mu,T)$-plane 
reaches all the way to the temperature axis,
cf.~Fig.~\ref{schem}. For larger masses the critical point moves to positive values
of $\mu$, while for smaller quark masses it is at imaginary values of $\mu$.
Our goal is to map out the functions $m_c(\mu), T^*(\mu)$ for imaginary $\mu$
and analytically continue their Taylor expansions to real $\mu$.

Our simulations were performed on $8^3,10^3$ and $12^3\times 4$ lattices. For each pair
$(m,\mu)$ we determined the critical coupling $\beta_c(m,\mu)$ by interpolating
between three $\beta$-values by means of Ferrenberg-Swendsen reweighting \cite{fs}.
For each $(m,\mu)$ we accumulated about $100k$ unit length 
trajectories generated by the R-algorithm
with step size $\delta\tau = 0.02$.
\begin{figure}[t]
\vspace*{3mm}
\centerline{\epsfxsize=7cm\hspace*{0cm}\epsfbox{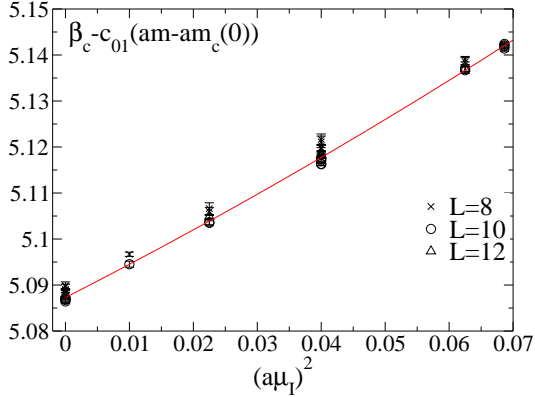}}
\vspace*{-0.8cm}
\caption[a]{
Combined pseudo-critical results for $L=8-12$ and various quark masses. Data for different $am$
are shifted to $a m_c(0)$ according to the
best fit, which is also shown.}          
\label{tc}
\end{figure}

\section{THE PSEUDO-CRITICAL SURFACE}

\begin{figure}[t]
\vspace*{3mm}
\centerline{\epsfxsize=7cm\hspace*{0cm}\epsfbox{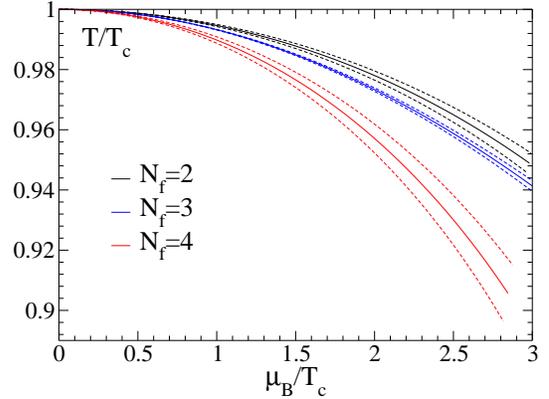}}
\vspace*{-0.8cm}
\caption[a]{
Pseudo-critical lines for different $N_f$.}          
\label{tc_nf}
\end{figure}

Data points for all three volumes are shown in \fig \ref{tc}. They are fitted
by truncated Taylor series 
\be
\beta_c(a\mu,am)=\sum_{k,l=0} c_{kl}\, (a\mu)^{2k}\, (am-am_c(0))^l.
\ee
Our data are accurate enough to be sensitive to next-to-leading terms. A 
$\chi^2$-analysis comparing all possible fits shows clear preference for a $\mu^4$-term,
while $m^2$- and mixed $m\mu^2$-terms appear to be smaller than our present errors.
On the other hand, fits on different volumes are consistent with each other, 
so we may fit all volumes together.
The best fit shown in the plot corresponds to 
$\beta_c(a\mu_I,am) = 5.1453(2) + 1.780(16) (am - 0.0323)
+ 0.705(10) (a\mu_I)^2 + 1.46(15) (a\mu_I)^4$.
Using the two-loop beta-function, this result is easily converted to continuum units,
\ba 
\frac{T_c(\mu,m)}{T_c(0,m_c(0))}&=& 1
+ 1.937(17) \left(\frac{m-m_c(0)}{\pi T_c}\right)\nn\\ &&\hspace*{-2cm} 
-0.602(9)\left(\frac{\mu}{\pi T_c}\right)^2
+0.23(9)\left(\frac{\mu}{\pi T_c}\right)^4.
\label{eqtc}
\ea
This curve is displayed in \fig \ref{tc_nf} together with earlier results
for two \cite{fp1} and four \cite{el} flavors. Note, however, that the present result
is the only one including a $\mu^4$-term. Our three flavor  result also
coincides within its small errors with that for $N_f=2+1$ presented in \cite{fk2}.
This is not surprising: $T_c(\mu)/T_c(0)$ should depend
little on the quark masses, as long as these are small compared to $\pi T$,
as is the case in \cite{fk2}.

\section{THE LINE OF CRITICAL POINTS}

\begin{figure}[t]
\vspace*{3mm}
\centerline{\epsfxsize=7cm\hspace*{0cm}\epsfbox{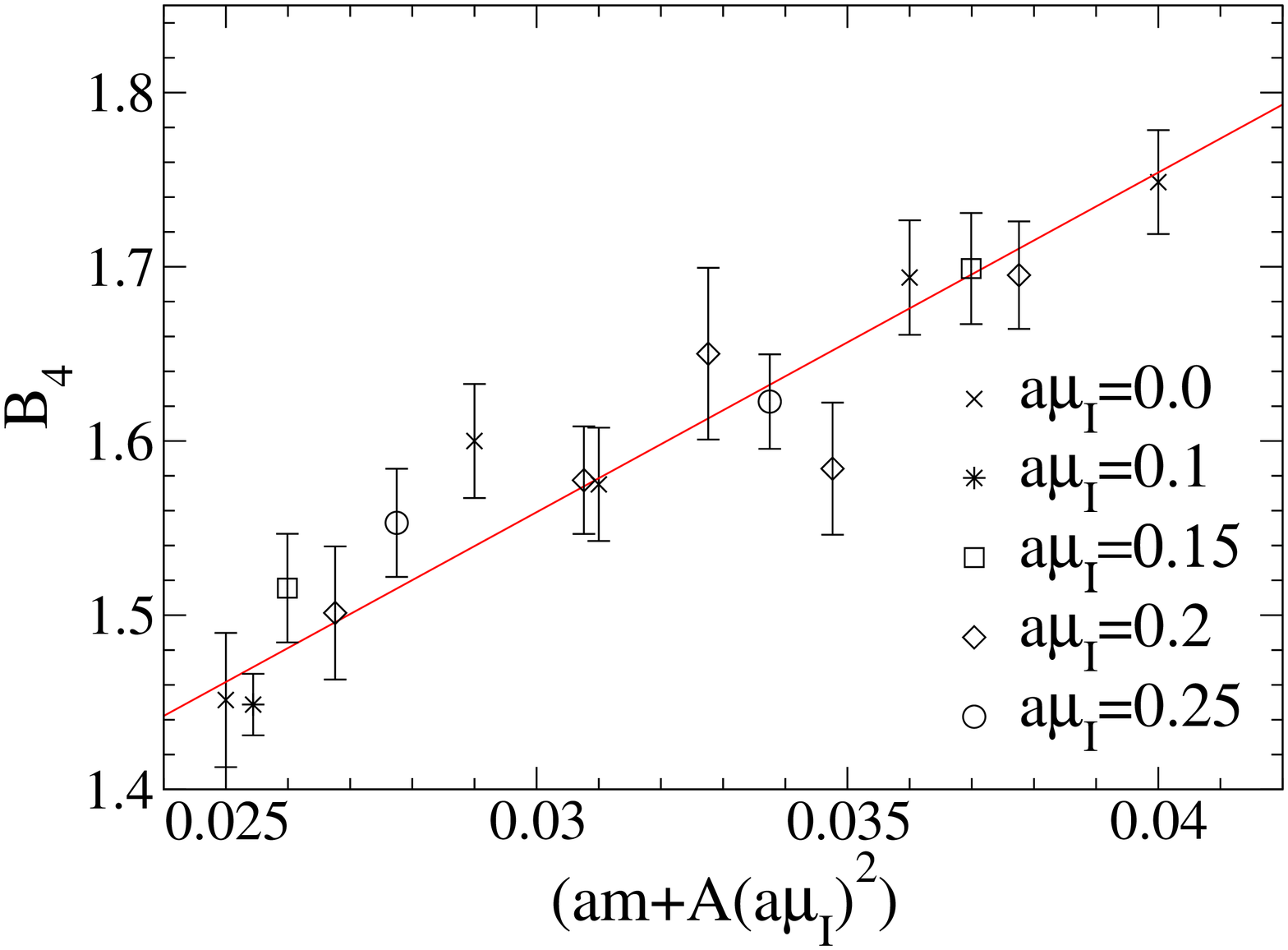}}
\vspace*{-0.8cm}
\caption[a]{
Fit to measurements of the Binder cumulant on $8^3$.}          
\label{B4_8}
\end{figure}

In order to locate the critical point we used the 
Binder cumulant of the chiral condensate,
\be
B_4=\frac{\langle(\delta\bar{\psi}\psi)^4\rangle}
{\langle(\delta\bar{\psi}\psi)^2\rangle^2},
\ee
measured across the pseudo-critical surface.
In the infinite volume limit, this quantity assumes a universal value
at a critical point. In the universality class of the 3d Ising model it is 
$B_4(m_c(\mu),\mu)=1.604$,
while the value is smaller for first order transitions and larger for crossover.
Like for the critical coupling, we fit our data to a Taylor expansion
\be
B_4=1.604 + B\left(am-am_c(0) + A(a\mu_I)^2\right).
\ee
Written in this form, $A$ yields directly the desired result
$A=d(am_c)/d(a\mu)^2$. We may now fit $15$ $(m,\mu)$-pairs measured on our $8^3$ lattice
by a single three-parameter fit, and find $A=0.044(19)$. Data and fit result are shown 
in \fig \ref{B4_8}.

\begin{figure}[t]
\vspace*{3mm}
\centerline{\epsfxsize=7cm\hspace*{0cm}\epsfbox{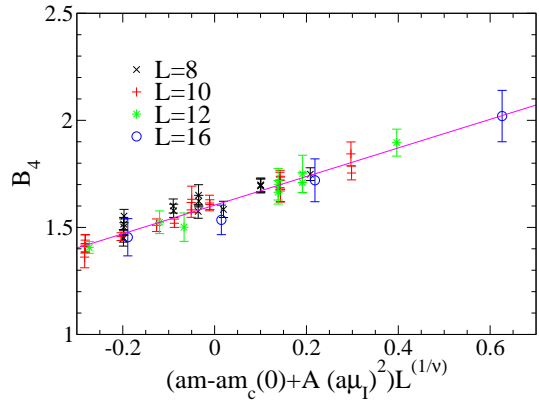}}
\vspace*{-0.8cm}
\caption[a]{
Fit to measurements of the Binder cumulant on all volumes ($L=16$ from \cite{mc0}).}          
\label{B4_all}
\vspace*{-0.4cm}
\end{figure}

Unfortunately, the corresponding data for $10^3,12^3$ are not accurate enough to
constrain the parameter $A$, which is consistent with zero on those volumes.
However, good fits for $B$ are obtained, and those can be used to check consistency with
universality. The latter predicts that, for $\beta=\beta_c(m,\mu)$, the correlation
length scales as $\xi \propto |m - m_c(\mu)|^{-\nu}$, and hence 
$B_4\left((L/\xi)^{1/\nu}\right)=B_4\left(L^{1/\nu}(am-am_c(\mu)\right)$.
We have fitted the volume behavior by $B\sim L^{-1/\nu}$, and obtain $\nu=0.62(3)$ in accord
with the 3d Ising value $\nu=0.63$. Having confirmed 
universal finite volume scaling for our data,
we may plot and fit all volumes together as in \fig \ref{B4_all}.
The figure nicely demonstrates the consistency of our results with 
universal finite size scaling.
The fit yields the value $m_c(0)=0.0323(3)$,
in perfect agreement with earlier studies done at zero density \cite{mc0}.
However, as we mentioned before, the data on the larger volumes do not constrain 
the parameter $A$ yet, so that we quote our $8^3$ number as the final result.
Converted to continuum units, this yields for the $\mu$-dependence of the critical
bare quark mass
\be
\frac{m_c(\mu)}{m_c(\mu=0)}=1 + 0.84(36) \left(\frac{\mu}{\pi T}\right)^2.
\ee
This coefficient is inconsistent with a preliminary one obtained
with improved 
actions \cite{schmidt}. Action-dependent multiplicative 
renormalization drops out of the 
ratio of quark masses, and any difference should be at most $\op(a^2)$. 
Moreover, the result \cite{schmidt} violates an analytic upper bound of 9 (see
\cite{fp2}). We stress the difficulty of measuring the Binder cumulant 
accurately, without underestimating the error: the tunneling frequency is low, 
and very long MC runs are necessary.

\end{document}